\documentclass{PoS}

\newcommand\araa{ARA\&A}%
\newcommand\apj{ApJ}%
\newcommand\apjl{ApJ}%
\newcommand\aap{A\&A}%
\newcommand\mnras{MNRAS}%
\newcommand\pasp{PASP}%
\def\lsim{\!\!\!\phantom{\le}\smash{\buildrel{}\over
 {\lower2.5dd\hbox{$\buildrel{\lower2dd\hbox{$\displaystyle<$}}\over
                                 \sim$}}}\,\,}
\def\gsim{\!\!\!\phantom{\ge}\smash{\buildrel{}\over
{\lower2.5dd\hbox{$\buildrel{\lower2dd\hbox{$\displaystyle>$}}\over
                               \sim$}}}\,\,}
\def\msun{\hbox{M$_{\odot}$}}
\def\kms{\mbox{\,km~s$^{-1}$}}
\def\mdot{\dot M}
\def\Mdot{\dot M}
\def\msunyr{\mbox{\,${\rm M_{\odot}\, yr^{-1}}$}}

\def\ergshz{\mbox{~erg~s$^{-1}$~Hz$^{-1}$}}
\def\arcsec{\ifmmode ^{\prime\prime}\else$^{\prime\prime}$\fi}

\title{Core-collapse and Type Ia supernovae with the SKA}
\ShortTitle{Core-collapse and Type Ia supernovae with the SKA}

\usepackage[authoryear,sort]{natbib}
\bibpunct{(}{)}{;}{a}{}{,}

\author{\speaker{M.A. P\'erez-Torres}$^{1,2,3}$, 
 A. Alberdi$^1$, R. J. Beswick$^4$, P. Lundqvist$^5$, 
 R. Herrero-Illana$^1$, C. Romero-Ca\~nizales$^{6,7}$, S. Ryder$^8$, M. della Valle $^{9}$,
 J. Conway$^{10}$, J.M. Marcaide$^{11,12}$, S. Mattila$^{13}$, T. Murphy$^{14}$, E. Ros$^{15,11,16}$  \\
$^1$Instituto de Astrof\'{\i}sica de Andaluc\'{\i}a (IAA-CSIC), E-18008 Granada, Spain\\
$^2$Centro de Estudios de la F\'{\i}sica del Cosmos de Arag\'on, E-44001 Teruel, Spain\\
$^3$Depto. de F\'{\i}sica Te\'orica, Facultad de Ciencias, Universidad de Zaragoza, E-50009, Spain\\
$^4$Jodrell Bank Centre for Astrophysics/{\it e-}MERLIN, The University of Manchester, M13~9PL, UK\\
$^5$Stockholms Observatorium, Sweden \\
$^6$Instituto de Astrof\'{\i}sica,  Pontificia
Universidad Cat\'olica de Chile, Macul, Santiago, Chile \\
$^7$Millennium Institute of Astrophysics, Vicu\~na Mackenna 4860, 7820436
Macul, Santiago, Chile\\
$^8$Australian Astronomical Observatory, P.O. Box 915 North Ryde NSW 1670, Australia \\
$^{9}$Osservatorio Astronomico di Capodimonte - INAF, 80131 Napoli, Italy\\
$^{10}$Onsala Space Observatory, Chalmers, Tekniska H\"ogskola, SE-439 92 Onsala, Sweden\\
$^{11}$Dept. Astronomia i Astrof\`{\i}sica, Universitat de Val\`encia, E-46100 Burjassot, Spain \\
$^{12}$DIPC, Paseo Manuel Lardizabal, Donostia-San Sebasti\'an, Spain\\
$^{13}$Finnish Centre for Astronomy with ESO (FINCA), University of Turku, FI-20014, Finland\\
$^{14}$The University of Sydney, 44 Rosehill Street, NSW 2006, Australia \\
$^{15}$Max-Planck-Institut f\"ur Radioastronomie, Auf dem H\"ugel 69, D-53121 Bonn, Germany\\
$^{16}$Observatori Astron\`omic, Universitat de Val\`encia, E-46980 Paterna, Val\'encia, Spain\\
E-mail: \email{torres@iaa.es}
}

\abstract{
{\bf Core-collapse SNe (CCSNe):}  
Systematic searches of radio emission from  CCSNe are still lacking, and only targeted searches of radio emission from just some of the optically discovered CCSNe in the local universe have been carried out. 
Optical searches miss a significant fraction of CCSNe  due to dust obscuration; CCSN radio searches are thus more promising for yielding the complete, unobscured star-formation rates in the local universe.  
The SKA yields the possibility to piggyback for free in this area of research by carrying out commensal, wide-field, blind transient survey observations. 
SKA1-SUR should be able to discover several hundreds of CCSNe in just one year, compared to about a dozen 
CCSNe that the VLASS would be able to detect in one year, at most. 
SKA, with an expected sensitivity ten times that of SKA1, is expected to detect 
CCSNe in the local Universe by the thousands. Therefore, commensal SKA observations could easily result in an essentially complete census of all CCSNe in the local universe, thus yielding an accurate determination of the volumetric CCSN rate. 
{\bf Type Ia SNe:} We advocate for the use of the SKA to search for the putative prompt ($\sim$first few days after the explosion) radio emission of any nearby type Ia SN, via target-of-opportunity observations. The huge improvement in sensitivity of the SKA with respect to its predecessors will allow to unambiguously discern which  progenitor scenario (single-degenerate vs. double-degenerate) applies to them.
}

\FullConference{Advancing Astrophysics with the Square Kilometre Array,\\
		June 8-13, 2014\\
		Giardini Naxos, Sicily, Italy }

\begin{document}

\section{Why CCSN searches in the radio?}
\label{sec:rsn-searches}


The limited sensitivity of pre-eMERLIN/VLA interferometric arrays has  biased
past radio observations of CCSNe towards the brightest events, preventing any
systematic radio follow-up of CCSNe of all types. The exception has been that of Type Ib/c SNe, which due to their GRB link have been the subject of a  systematic monitoring with the VLA \citep[see, e.g.,][]{soderberg06,bietenholz14}.  All this makes the currently existing radio observations of
CCSNe of rather limited use 


In this chapter, we argue that a commensal, wide-field, transient survey with the SKA could potentially allow us to obtain a complete census of CCSNe in the local universe, and therefore will permit us to determine the true CCSN rate and thus the star-formation rate of the population of massive stars in the local universe. 
 In addition, some specific, relevant questions that will be tackled by those observations include the following:

\begin{itemize}

  \item {\it Obtaining an accurate estimate of the true volumetric CCSN rate in the local universe}, $\Re$. 
   Wide-field SKA observations covering a significant  area of the sky will discover  many CCSNe in the nearby universe, and therefore will allow us to determine this relevant parameter (see Fig.~\ref{missing-ccsne}).

	\item {\it Probing the   SN-CSM interaction for all CCSN types}, from the
  relatively faint Type IIP to the extremely radio bright Type IIn
  SNe (Fig.\ref{Lpeak}). Probing the SN-CSM interaction for all CCSN types will allow us to obtain basic,
crucial information to characterize their progenitors, including
mass-loss rates and, for synchrotron-self-absorbed SNe, the shock
radius and the magnetic field--directly from the light curves (see, e.g.,  \citealt{chevalier98}).
  
 \item \emph{Bridging the gap between Type Ibc SNe and (long) $\gamma$-ray bursts}.  Type Ibc are arguably the CCSNe that show the highest blastwave speeds, yet most of them are energetically much less powerful than GRBs. Recently, however, cases like SN 2009bb, with $\beta \sim 0.9$ and energy $\sim 10^{49}$~erg seem to be intermediate cases. These "engine-driven" CCSNe  could be detected
with the high-sensitivity offered by the SKA, thus 
filling this gap in the energy-blastwave velocity parameter space of SNe-GRBs 
(see, e.g., \cite{gal-yam06}  and references therein).
   
  \item \emph{Typing CCSNe from their radio behaviour.}
  A systematic monitoring could permit us to type CCSNe from their radio
emitting properties.  This is crucial for the study
of the hidden SN population in (Ultra)Luminous Infra-Red Galaxies, where a spectroscopical
classification, or even an optical discovery, is essentially impossible 
(see, e.g., \cite{mattila12} and Fig.\ref{missing-ccsne}).

  \item \emph{Unveiling the hidden CCSN population.}
\citet{horiuchi11} found out  an apparent mismatch between the measured CCSN rate (mostly from optical observations) and the cosmic massive star formation rate, which was twice as large as the measured one.
However, \citet{mannucci07} 
and \citet{mattila12} have shown that a significant fraction of the exploding CCSNe in the local universe are hidden behind dust, and that this fraction seems to increase significantly as one goes back in the history of the universe (see Fig. \ref{missing-ccsne}, left panel), i.e., there seems to be no mismatch when the 
\emph{hidden CCSNe} are taken into account. The results obtained by \citet{mattila12} were mostly based on high spatial resolution near-IR and radio observations of the local LIRG Arp 299, and showed that the combination of near-IR and radio observations with high angular resolution are very useful to study CCSNe in LIRGs (e.g. \citealt{kankare14,crc14}). 
Radio observations have the advantage over both optical and near-IR that the emission from CCSNe is not hampered by dust, and thus offers an excellent opportunity to determine the true core-collapse supernova rate in the local universe. In fact, most of the the CCSNe that explode in the compact, central regions of luminous and ultra-luminous infrared galaxies can essentially be found only at radio wavelengths, e.g. SN~2000ft in NGC 7469 \citep{colina01,alberdi06}, or the many CCSNe unveiled by radio observations in Arp 299A \citep{mpt09a,bondi12},  Arp 299B \citep{crc11}, or Arp 220 \citep{parra07,batejat11}. 
   
\begin{figure}	
  \begin{center}
  \includegraphics[width=\textwidth]{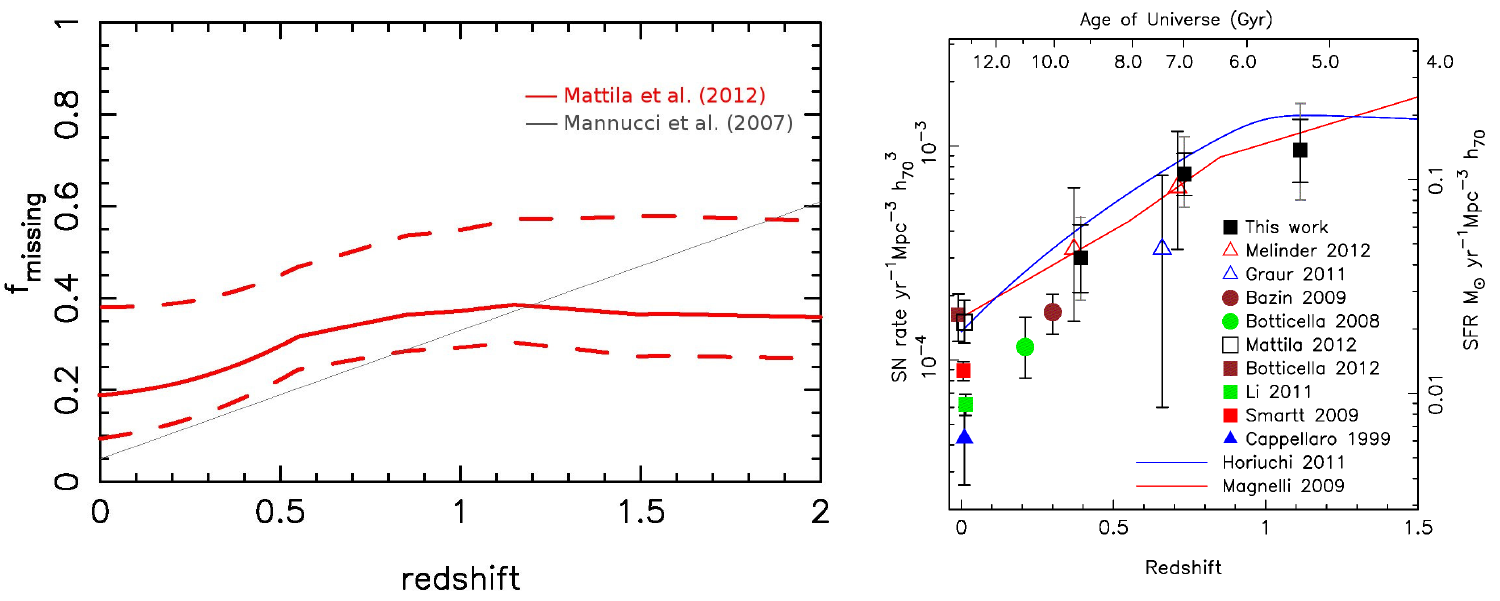}
  \caption{\textit{Left:} Fraction of CCSNe missed by rest-frame optical searches as a function of redshift. The red lines show the best estimate (solid line) together with the upper and lower bounds 
of the missing fraction as dashed lines. The solid black line corresponds to the missing
fraction from \citet{mannucci07}. (Figure from \citealt{mattila12}.) 
\textit{Right:} Core-Collapse Supernova rate as a function of redshift. (Figure from \citealt{dahlen12}.)
The use of the SKA as a CCSN discovery machine willl  reduce the relatively large uncertainty in the missing fraction of CCSNe in the local universe (left figure), as well as in the volumetric CCSN rate (right figure).
	}
			\label{missing-ccsne}
		\end{center}
	\end{figure}

  \item \emph{Correlating optical and radio properties.} The combined use of optical information for both SN and host galaxy, together with the obtained peak (radio) luminosities will allow us to check whether there is a correlation between the optical and radio properties of CCSNe, as well as  with their host galaxies. This will be possible by, e.g., making a combined, commensal, use of wide-field surveys programmed at radio wavelengths with SKA, and at optical wavelengths with, e.g., the LSST or similar telescopes. Obviously, the most interesting cases will likely be subject of targeted, monitoring observations with these and other facilities.
For example, \cite{gal-yam06} clearly showed the impact of carrying radio and optical follow-up observations of all possible radio transients discovered in surveys covering a significant fraction of the sky area, in terms of GRBs and SNe studies.

\end{itemize}

\section{Radio emission from CCSNe}
\label{sec:rsne}

In this section we give account of the radio emitting properties, which are crucial to understand the technical issues discussed in the remaining sections.

 CCSNe result from both single and binary massive star systems, with initial masses $\gsim8$ \msun, and include a diversity of spectroscopically divided subtypes (Type IIP/L, IIn, IIb, Ib/c; see e.g. \citealt{eldridge13}).  Upon gravitational collapse of the core, the outer parts of the shell are ejected at very high speeds, $v \approx 10000\, (E_{\rm in}/10^{51} {\rm erg})^{1/2} \, (M_{\rm ej }/1 \msun)^{-1/2}\kms $,
and the interaction with the outer, much less dense, material of the progenitor star, drives a blastwave that propagates at speeds as high as $v_s \simeq (0.1-0.3) c$ \cite[e.g.,][]{matzner99}. This high-speed shock heats the circumstellar material to temperatures of $\sim\,10^9$\,K.  
In addition, a reverse shock propagates back into the stellar envelope at speeds of
500-1000 ${\rm km\,s^{-1}}$ relative to the expanding ejecta.

 When the supernova shock-wave ploughs through the circumstellar gas, a high-energy density shell forms. Within this shell, electrons are accelerated to relativistic speeds and significant magnetic fields are generated, giving rise to the production of the observed  radio emission. 
The radio emission from supernovae is essentially non-thermal synchrotron emission from relativistic electrons, and  is due to circumstellar interaction \citep[e.g.,][]{chevalier82}. 
In the framework of the SN-circumstellar interaction scenario, the optically thin radio emission from CCSNe scales with the mass-loss wind parameter, 
$\mathcal{M} = \mdot / v_w$, 
where $\mdot$ and $v_w$ are the pre-supernova mass-loss rate  and wind speed, respectively. 
Thus, for the radio emission  to be appreciable, the progenitor star must have lost a significant amount of mass via a stellar wind and/or stripping due to a close companion. The presupernova stellar wind speed plays an equally important role:  the smaller the wind speed, the more mass is retained in the vicinities of the SN progenitor, and the stronger the circumstellar interaction will be. 
According to their $\mathcal{M}$ value, CCSNe can be divided into two basic groups: 
Type Ib/c SNe, which  have expelled fast winds ($v_w \approx 1000$ \kms), have $\mathcal{M} \sim 10^{-8} \msunyr/\kms$, thus implying $\mdot \sim 10^{-5} \msunyr$  during their blue supergiant phase.
Type II SNe are characterized by  significantly slower ($v_w \approx 10$ \kms) winds, although the large differences in mass-loss rates $\Mdot \sim (10^{-4}-10^{-6}) \msunyr$ result in a large scatter in 
$\mathcal{M} \sim (10^{-5} - 10^{-7}) \msunyr/\kms$ during their red supergiant phase, 
which affects the time at which they reach their peak radio luminosity.

This pre-supernova wind, made of thermal electrons, has a power-law density profile, which for a steady, spherically symmetric wind has the following form:
$\rho_{\rm CSM} \propto r^{-2}$. 
This circumstellar wind is the main responsible for the partial suppression of 
synchrotron radio emission from supernovae (e.g., \citealt{chevalier82}), so that
the radio flux density evolution of many CCSNe can be well described by the following equation \citep[e.g.,][]{weiler02}

\begin{equation}
		S_\nu (t) = K_1 \left( \frac{\nu}{\rm 5 GHz} \right)^\alpha
					\left( \frac{t - t_0}{\rm 1 day} \right)^\beta e^{-\tau_{\nu}},
\label{eq:Snu}
\end{equation}

\noindent 
where 
$\tau_{\nu}$ is the optical depth due to external absorption, which for
the sake of simplicity is taken to be due only to the (thermal) electrons of the pre-supernova wind and of a possible distant H~II region. 
The optical depth can then  be written as

	\begin{equation}
		\tau_{\nu} =  \left( \frac{\nu}{\rm 5 GHz} \right)^{-2.1}
		                       \left[ K_2\ \left(\frac{t - t_0}{\rm 1 day} \right)^{\delta}\ + K_3 \right],	
	\label{eq:tau}
	\end{equation}

\noindent 
where the first term accounts for the time varying external absorption, which continuously decreases (since $\rho_{\rm CSM} \propto r^{-2}$), and the second term allows for a non-varying absorber, i.e., a distant H~II region. The former term is usually the one that dominates in most CCSNe, but cases where the latter term dominates also exist, 
 e.g., SN~2000ft in NGC 7469 \citep{alberdi06,mpt09b}, or some of the SNe in Arp 299A \citep{mpt09a,mpt10}.
(Synchrotron self-absorption seems to be the dominant mechanism in many Type Ib/c SNe; however, this does not affect the discussion below, so for the sake of simplicity we only consider here external absorbers.)

In Table \ref{rsne-params}  we present typical values for the relevant radio emission parameters 
in Eqs. \ref{eq:Snu} and \ref{eq:tau},  based on those published in \cite{weiler02} and \cite{crc14} for a sample of CCSNe.
Stripped-envelope SNe (i.e., Type Ib/c and, to some extent, their cousins, Type IIb SNe) have an optically thin spectral index, $\alpha$ ($S_\nu \propto \nu^\alpha$), which is steeper than the spectral index of Type II SNe. The combination of high blastwave speed and stripped envelope in Type Ib/c events 
(and also in Type IIbc supernovae) explains their very short times to peak and their swift radio evolution (steep values of $\beta$), compared to the rest of CCSNe. 

\begin{figure}	
		\begin{center}
			\includegraphics[width=16cm]{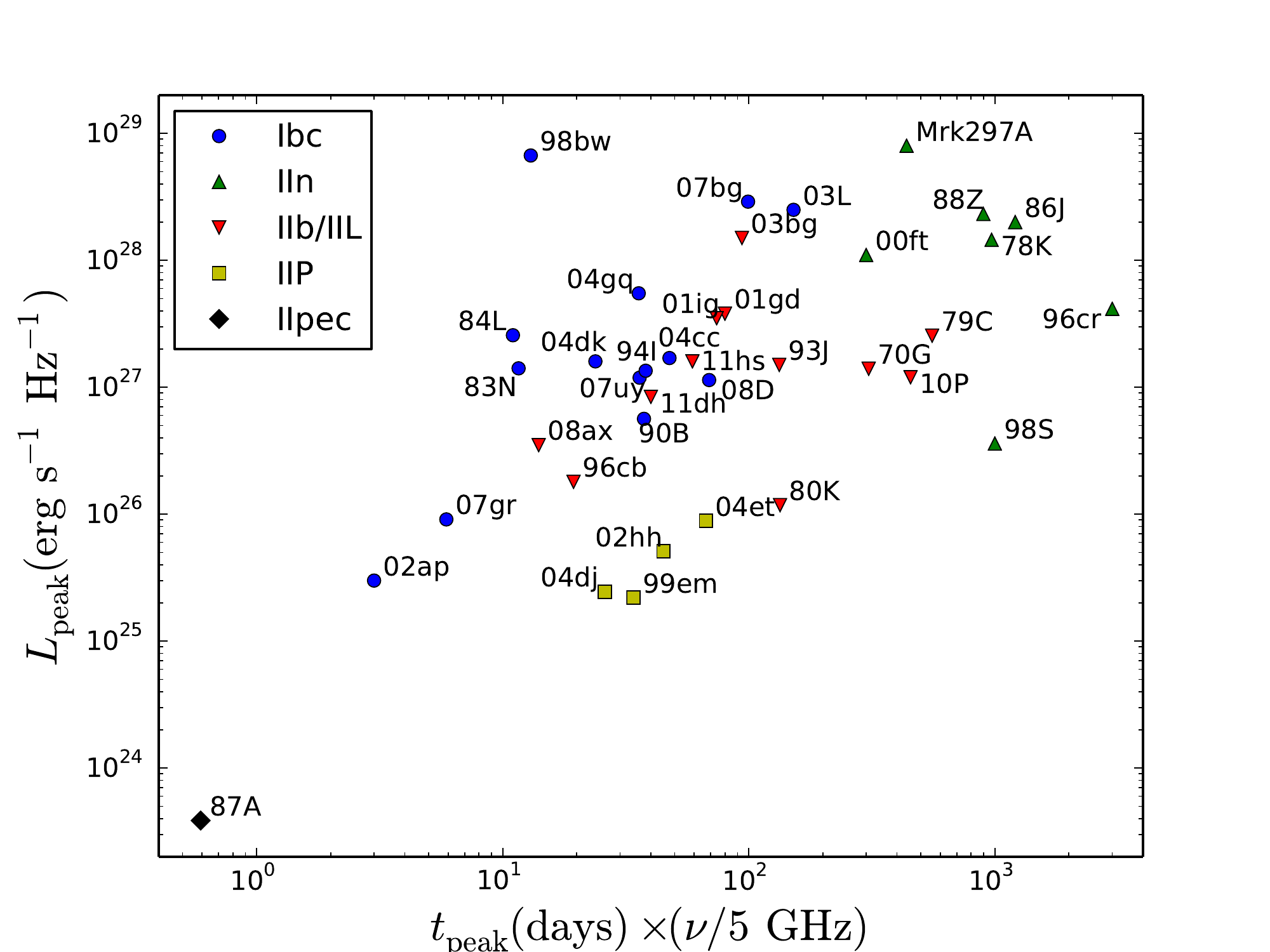}
			\caption{ \em Peak spectral radio luminosity of CCSNe 
			vs. time to peak,  times the observing frequency.  
There is a clear correlation between  time to the peak and the luminosity at the peak;
in this parameter space, each subtype of CCSNe are located in specific areas;
 CCSNe show typical peak radio luminosities above  $10^{26}$\ergshz, except  
Type IIP SNe, which usually peak at a few times $10^{25}$\ergshz, and the very rare 87A-like events. 
Figure adapted from \cite{chevalier06} and \cite{crc14}.
	}
			\label{Lpeak}
		\end{center}
\end{figure}

The tabulated fraction of each CCSN type is taken from \cite{eldridge13}, 
which used CCSNe discovered between 1998.00 and 2012.25 (14.25yr) in galaxies with recessional velocities less than 2000 \kms.
Type Ibc SNe  make up $\sim26\%$ of all CCSNe in the local 
universe. 
$\sim$57\% of all CCSNe are faint radio SNe, i.e., the Type IIP SNe and the peculiar Type II SNe, e.g., SN 1987A; 
$\sim$15\% are the moderately bright Type IIb and Type IIL SNe, and only $\lsim$3\% are type IIn SNe, the brightest radio events (see Table \ref{rsne-params} and Figure \ref{Lpeak}). 

Type IIn SNe show peak luminosities  $L_{\rm peak} \gtrsim (1 - 2) \times 10^{28}$ \ergshz, 
while Type Ib/c and Type IIb SNe peak at values of a few times $10^{27}$ \ergshz.  
Together, these bright CCSNe account for less than 30\% of all radio supernovae. 
Most of the remaining $\sim$70\% are type IIP SNe, which are significantly fainter radio supernovae, with peak luminosities $\lesssim$ few times $10^{26}$ \ergshz (see Fig. \ref{Lpeak}).

\begin{table*}
\caption{Radio parameters of core-collapse  supernovae}
\scalebox{0.9}{
\begin{tabular}{lrcccrr}
\hline
SN Type & $f_{\rm CCSN}$ & $\alpha$ & $\beta$ & $\delta$  &
                 $t_{\rm peak (days)}\, \times (\nu/5\ {\rm GHz})$ & $L_{\nu, {\rm peak}}/ (10^{27} {\rm erg\,s^{-1}\,Hz^{-1}}$)   \cr
\hline
 Ib/c  & 26.0\% & $-1.1$ & $-1.4$  & $-2.5$ & 2 - 100 & 0.02 -- 20   \cr 
 IIb   & 12.1\% & $-1.1$ & $-1.0$ & $-2.0$ &  180 & 2 -- 5  \cr 
 IIP   & 55.5\% & $-0.7$ & $-$(0.7 - 1.2) & $-$3.0 &  30 -- 500 & 0.2 -- 1 \cr 
 IIL    & 3.0\% & $-0.7$ & $-0.8$ & $-$(2.7 - 3.0) & 100 -- 800 & 1 -- 30 \cr 
 IIn   & 2.4\%  & $-0.7$ & $-$(1.3 - 1.7) & $-$3.0 &  $\gsim 800$ &  10 -- 20 \cr 
 87A-like & 1.0\% & $-1.0$ & ... & $-$3.0 &  $\sim$2  &  0.004 \cr 
\hline
\end{tabular}
}
\label{rsne-params}
\end{table*}

\section{CCSN searches with the SKA}

In this section,  we discuss in detail the advantages of a commensal, wide-field, transient survey with the SKA with respect to currently envisioned efforts with other facilities.

In fact, several wide-field sky surveys will be carried out with the SKA pathfinders MeerKAT and ASKAP, 
as well as with the upgraded Very Large Array. Those surveys can be used commensally for transient studies, by profitting from programmed wide-field observations.  
For example, the planned Very Large Array Sky Survey (VLASS) is contemplating the possibility of 
observing wide field areas (a few hundreds to about one thousand of square degrees)
with nominal sensitivities of  $1 \sigma \simeq 100 \mu$Jy/beam per epoch, aiming at reaching r.m.s. values of $\simeq (40 - 70) \mu$Jy/beam after stacking multi-epoch observations. 
However, those  sensitivities are just too shallow to be of any real use for CCSN studies.
Indeed, a $5 \sigma$ figure of merit corresponds to 500 $\mu$Jy/beam per epoch, so that the maximum distance to detect a type IIP event would be $\sim$9 Mpc (see Table \ref{expectations}). Unless sky areas close to the full celestial sphere are surveyed--which is very unlikely-- VLASS will only pick up a handful of CCSNe after one year of observations (see Sect. \ref{sec:strategy} below). Most likely, this handful of CCSNe will be discovered first by optical searches, and some of them will be subject of targeted radio observations anyway, given their vicinity. 
Thus, deeper sensitivies are needed to make a substantial contribution to the field.

\subsection{SKA survey strategy for commensal CCSN searches}
\label{sec:strategy}

The best strategy for transient studies with the SKA is one that combines good angular resolution ($\lesssim 1.5$ arc sec) and large field of view, (FoV $\gtrsim$10 deg$^2$), at frequencies around and above $\sim$1.5 GHz. 
The SKA1-SUR, which should provide an approximate survey sensitivity of 
$\lesssim 4.2 \mu$Jy/beam after 1-hr of on-source integration (assuming a bandwidth of 500 MHz) for a FoV of $\simeq$18 deg$^2$, with an angular resolution of $\sim$0.9 arc sec at a fiducial frequency of 1.7 GHz, meets such requirements (see Table \ref{ska-params}).

SKA1-SUR is likely to observe the sky for $\gsim$1000 hr in its first year of operations. Thus, depending on the specific observing band (or bands), the surveyed area will lie in the range of $\sim$9000 $-$ 18000 deg$^2$. 
As we explain below, to be of use for CCSNe searches, each field of view should be visited five times (each time for an on-source time of 12 minutes), at a cadence of one visit every $\lesssim 90$d. 
The r.m.s.\, attained per pointing will be of 9.3 $\mu$Jy/beam per pointing, and the r.m.s. for stacked images will be 4.2 $\mu$Jy/beam, according to the SKA1-SUR
specifications.
Such a commensal strategy is likely to result in the largest sample of radio supernovae ever detected in the nearby universe. 

\begin{table*}
\caption{Relevant observational parameters for the proposed SKA1 designs and comparable telescopes}
\scalebox{0.9}{
\begin{tabular}{lccccccc}
\hline
    & eMERLIN & VLA$^\dagger$  & MeerKAT & ASKAP & SKA1-SUR & SKA1-LOW & SKA1-MID  \cr
\hline
FoV (deg$^2$)   & 0.25 & 0.25  & 0.86  & 30  & 18 & 27  & 0.49 \cr
Fiducial Freq. (GHz)  & 1.4 & 1.4 & 1.4 & 1.4 & 1.67 & 0.11 & 1.67 \cr
Resolution (arcsec)   & 0.15  & 1.4 & 11 & 7  & 0.9  & 11 & 0.22 \cr
Baseline/size  (km) & 220 &   35 & 4 & 6 & 50 & 50  & 200 \cr
Bandwidth (MHz) & 400 & 1000 & 1000 & 300 & 500 & 250 & 770  \cr
Sensitivity ($\mu$Jy-hr$^{-1/2}$)    & 27.1  & 3.9 & 3.2 &  28.9 & 3.7 & 2.1 & 0.7\cr
\hline
\end{tabular}
}
$^\dagger$The VLASS 1-$\sigma$ figure is $\gsim$10 times worse than obtained with one hr of VLA observing time.
\label{ska-params}
\end{table*}

\subsection{Expectations for CCSN discoveries with wide surveys}

Table \ref{expectations} summarizes the expectations for detecting CCSNe using the VLASS, SKA1-SUR, SKA-50\% and SKA. More specifically, we show the maximum distance to which a CCSN is expected to be detected above 5$\sigma$, for the nominal r.m.s. values of VLASS ($\sim 100 \mu$Jy/beam/pointing) and SKA1-SUR ($\sim 9.3\mu$Jy/beam/pointing), as well as for SKA1-50\% (i.e., only half of the bandwidth of SKA1-SUR will be available), and SKA (ten times more sensitive than SKA1).
Here, we took the approximate median value of the peak luminosity of each CCSN type. In practice, this means that only half of the quoted values in Table \ref{expectations} will be detected. The other half would lie below our 5~$\sigma$ cut and are expected to be missed.
In its currently assumed specifications, VLASS will detect 
Type Ib/c SNe only up to a distance of $\lesssim$58 Mpc, less than half the value quoted in \cite{kamble14}.
SKA1-SUR, on the other hand, will be able to detect Type Ib/c SNe up to $\sim 188$ Mpc, i.e.,
a volume $\sim$35 times larger than the volume where the VLASS will be sensitive to Type Ib/c SNe.

For an all-sky survey, the total number of detections in one year can be written as  
$N_{\rm all-sky} = \Re \times V \times {\rm min[\Delta t_{\rm bright} / \Delta t_{\rm cadence}, 1]}$, where $\Re$ is the  volumetric rate of CCSNe,  $\Delta t_{\rm bright}$ is  the time that a CCSN of a given type remains bright above the survey sensitivity, and   $\Delta t_{\rm cadence}$ is the cadence time of the observations, respectively. As long as $\Delta t_{\rm bright}  \gtrsim \Delta t_{\rm cadence}$, one can be sure that all CSSNe are detected. Since 
$\Delta t_{\rm bright} \simeq \Delta\,t_{\rm peak}\, \nu_5^{-1}$ days (see col. 6 in Table \ref{expectations}),  then
$\Delta t_{\rm cadence} \lesssim \Delta\,t_{\rm peak}\, \nu_5^{-1}$ days, 
where $\nu_5 = \nu / 5 {\rm GHz}$. 
The volumetric rate of CCSNe in the local universe ($z \lsim 0.1$) is $\Re \sim 10^{-4}$ SNe yr$^{-1}$ Mpc$^{-3}$ within a factor of $\sim$2, while at $z \gtrsim 1.0$ the volumetric CCSN rate is $\Re \sim 10^{-3}$ SNe yr$^{-1}$ Mpc$^{-3}$ 
(see \citealt{dahlen12} and the right panel of Fig. \ref{missing-ccsne}).
The total number of CCSNe expected to be detected in one year over the whole sky, in the local universe ($z \lsim 0.1$), where $\Re \sim 10^{-4}$ SNe yr$^{-1}$ Mpc$^{-3}$ is then
$ N_{\rm all-sky} \approx 420\, \left(D_{\rm max}/100\, {\rm Mpc}\right)^3\, \left( \Re /10^{-4} {\rm SNe\, yr^{-1} Mpc^{-3}} \right)$, 
and the expected number of detections of each CCSN type is
$f_{\rm CCSN} \times N_{\rm all-sky}$,  
where $f_{\rm CCSN}$ is the fraction of each CCSN type (see Table 1). 
The cadence time is set up by the need to detect the fast-evolving type Ib/c SNe. For a fiducial frequency of 1.7 GHz, the cadence time should thus be no more than $\sim$90 days. 
Note the huge difference in the number of expected CCSNe to be detected by the VLASS and by SKA1-SUR, which 
is due to the VLASS being $\simeq$10.8 times less sensitive than SKA1-SUR. Since   
$N_{\rm det} \propto D_{\rm max}^3 \propto S_\nu^{-3/2}$, the expected number of detections in SKA1-SUR is $\simeq$ 35.3 times larger than for the VLASS.

We used the  above expression for $N_{\rm all-sky}$, and combined it with the fraction of SNe from each type (second column in Table 1) to obtain the expected number  of detected CCSNe  in one year,  for a  survey covering 10,000 deg$^2$ (see Table \ref{expectations}).
We also took into account the limiting detectable volumes for each CCSN type, according to their approximated median luminosities. The main limitation is due to the relatively small value of the maximum distance that will allow a
detection of a type IIP SN, which is why so few detections of them are expected.
(Note that the VLASS will need to sample a 10,000 deg$^2$ area for 25 years to blindly discover just one single type IIP event.)
On the contrary, the extremely bright Type IIn supernovae would be detectable, in principle, up to a distance of $\sim$422 Mpc ($z \sim 0.1$) with SKA1, and up to $\sim$1350 Mpc ($z \sim 0.25$) with SKA, assuming a tenfold increment in sensitivity (see Table \ref{expectations}).
The relatively high luminosity of spirals at those frequencies ($\sim 7\times10^{27}$\ergshz) may prevent  unambiguous detection of even Type IIn at those distances with SKA1, as the synthesized beam of 1.0\arcsec 
corresponds to 2.1 kpc, which could pick up a significant amount of the galaxy luminosity, thus the need for high angular resolution. This limitation will be overcome once SKA is completed, as is foreseen to have a twentyfold better angular resolution.
In addition, one has to take into account that around 10\% of the massive star-formation already at $z \sim 0.1$ will come from Luminous Infrared Galaxies \citep{magnelli11}. While those galaxies are prolific CCSN factories, their detection with SKA1, or even with SKA, will not be possible in general, as the compact starburst (size $\lsim$ 500 pc) will have a typical 1.4 GHz brightness well in excess of our 5-$\sigma$ sensitivity limit. 
For the sake of simplicity, and to stay on the conservative side, we assume here that we recover all Type IIn SNe up to an effective distance of 250 Mpc with SKA1-SUR and SKA1-50\% (see Table \ref{expectations}).

\begin{table*}
\caption{Expectations for CCSN detections in the local Universe from commensal radio surveys from the
VLASS (5-$\sigma$ = 500 $\mu$Jy/beam), SKA1-SUR (5-$\sigma$ = 46.5 $\mu$Jy/beam), 
SKA1-50\% (5-$\sigma$ = 66.4 $\mu$Jy/beam), and SKA (5-$\sigma$ = 4.65 $\mu$Jy/beam)
assuming each survey observes at a nominal frequency of 1.7 GHz and covers 
an area of 10,000 deg$^2$ in one year.
$L_{\nu, 26} = L_{\nu, {\rm peak}} / 10^{26}$ erg/s/Hz; $\nu_5^{-1} = \nu / 5$ GHz
}
\scalebox{0.9}{
\begin{tabular}{lccrlrlrlrr}
\hline
SN Type &  $\Delta\,t_{\rm peak}\, \nu_5^{-1}$ & $L_{\nu, {\rm 26}}$ &  \multicolumn{2}{c}{VLASS} & \multicolumn{2}{c}{SKA1-SUR} &
      \multicolumn{2}{c}{SKA1-50\%} & \multicolumn{2}{c}{SKA} \cr
 \cline{4-11}
        &    [days] &  &  D$_{\rm max}$  &  N$_{\rm det}$&  D$_{\rm max}$ & N$_{\rm det}$
        & D$_{\rm max}$ & N$_{\rm det}$ & D$_{\rm max}$ & N$_{\rm det}$  \cr
\hline
 Ib/c     &   30             &    20 &  58 &   5.1  & 189   & 177 & 159 & 106 & 596 & 5618\cr 
 IIb, IIL &   $\sim$150 &    10 &  41 & 0.8   &  133   &  29  & 112.2 & 17.4   & 422 & 924\cr  
 IIP       &   40             &    0.5   &  9  &  0.04 &     30   &  1.5      & 25 & 0.9        & 94   & 47\cr 
 IIn       &  1000          & 100  &  129 &  6.6 &  422  &  104 & 355 & 80 & 1334 & 7247\cr  
 87A    & 2                 & 0.04  &   2.6 &  $\sim10^{-5}$ & 8.4 & $\sim 10^{-3}$ & 9 & $\sim 10^{-3}$ & 26.7 & 0.05\cr
\hline
Total  &                       &           &       &  $\sim$13    &       & $\sim$311   &  & $\sim$204  &  &  
$\sim$13800 \cr
\hline
\end{tabular}
}
\label{expectations}
\end{table*}

We advocate for SKA1-SUR transient observations at 1.7 GHz, which should  
provide a survey speed of $\simeq$18 deg$^2$/hr.  Thus, 10,000 deg$^2$ can be covered in 556 hr, which easily accomodates into surveys like SKA1-SUR. 
Indeed, the only requirement for such a programme to be successful under SKA1-SUR is that a minimum number of visits are done to the same field to secure the discovery of recently exploded CCSNe by means of their variability.
The cadence time would be limited by peculiar Type II SNe, e.g., SN 1987A, peaking very early after the explosion, but since we will miss them anyway (see Table \ref{expectations}), the cadence time is actually constrained to $\sim 90$ days by the type Ib/c SNe turnover. This requirement can be easily fulfilled by carrying out five 
equally spaced 12-min visits of the same field during one year, resulting in an r.m.s. of 9.3 $\mu$Jy/beam per visit and thus reaching the specified 4.2$\mu$Jy/beam after five visits (=1 hr on-source).  

While the VLASS will detect only $\sim$13 CCSNe after surveying 10,000 deg$^2$ after one year, 
SKA1-SUR is expected to detect over 300 CCSNe in one year (see Table \ref{expectations}). 
SKA1-SUR, unlike VLASS, will not need to cover an extremely huge area to ensure a large number of detections.
In fact, Table \ref{expectations} shows that the VLASS  needs to cover an area of $\gsim$13,000 deg$^2$ to detect just one single IIb/IIL event in one year; the detection of a single Type IIP from a blind radio survey within the VLASS,  would require about six years of observations of the whole sky (or 25 years of the same 10,000 deg$^2$ field). In short, the handful of radio SNe expected to be detected by 
the VLASS will most likely be detected anyway by nearby optical searches. 
Clearly, VLASS-class blind radio surveys will not add much to our knowledge of CCSN radio properties and/or statistics. On the contrary, SKA1-SUR, carried out over $\sim 10000$ deg$^2$ with a cadence time of 90 days, should detect about 177 type Ibc SNe, 29 type IIb/IIL, $\sim$1-2 type IIP, and 104 type IIn CCSNe, for a total of  $\sim$311 CCSNe in just one year, or about 26 CCSNe per month, for a total on-source time of 556 hr.  
Assuming a 33\% overhead time, the total time needed is of $\sim$740 hr, or 2.0 hr/day for such a commensal, transient survey to be successful.

There are several reasons for the unrealistic predictions made by \cite{kamble14} regarding expectations of CCSN detections within the VLASS. 
\cite{kamble14} mimicked the approach of \cite{lien11}, who fitted the distribution of radio luminosity peaks of about 20 well observed radio supernovae, concluding that the average peak radio luminosity of \emph{all} CCSNe was $\gsim 10^{27}$\ergshz. However, as Fig.~\ref{Lpeak} 
shows, different types of CCSNe peak at significantly different luminosities, so the predictions of both \cite{lien11} and \cite{kamble14} were too optimistic. 
Indeed, \citet{kamble14} considered all CCSNe to be type Ib/c, i.e., $f_{\rm Ib/c} = f = 1.0$, while
 type Ib/c in the local Universe make up about 26\% ($f_{\rm Ib/c} = 0.26$) of all CCSNe. 
Finally, \cite{kamble14} assumed that CCSNe are detected at the 1-$\sigma$ level, 
which clearly cannot be the case. Actually, for a clear detection, a minimum threshold of five times the off-source r.m.s. will be needed, since the number of independent beams within the 1.7 GHz FoV will be of several millions with SKA1, and this number will be much larger with the SKA. Therefore, artificial sources above 3$\sigma$ and approaching a $\sim 5\sigma$ level should be expected, which may result in a substantial fraction of false detections even with a 3 $\sigma$ threshold.

When the characteristic radio luminosity of each CCSN type is taken into account, as well as their fraction, and a realistic 5-$\sigma$ detection threshold, numbers get dramatically lower for the intended VLASS in terms of (blind) CCSN radio discoveries. Even for SKA1-SUR, Achille's heel remains the expected low-number statistics for Type IIP SNe, so large areas need to be covered (see Table \ref{expectations}).

\subsection{Impact of a ``poor man's'' SKA1}

We now quantify the impact of a decrement by a 50\% in the SKA1-SUR specifications (SKA1 50\%).
If the angular resolution is degraded, this is expected to have a minor effect in terms of CCSN discovery, as most of the SNe in normal, spiral galaxies, will be far from the galaxy nucleus, and therefore should be easily discerned.
However, the impact due to a poorer array sensitivity is very relevant.
Since $N_{\rm det}  \propto S_\nu^{-3/2}$, then a decrement of the available bandwidth by 50\% implies a sensitivity decrement by $\sqrt{2}$, and thus a decrement in the 
 number of CCSN detections of almost 40\%, if the initial specifications are not observed. 

Although the preferred frequencies are $\gsim 1.7$ GHz, 
which optimizes FoV, sensitivity, and angular resolution, as well as permits 
a cadence time significantly larger than at higher frequencies, any of the currently deep, very wide continuum surveys will have a profound, extremely positive impact in the number of CCSN discoveries.
In particular, any of the
two All-Sky referenced surveys for SKA1 at band 2 will be very beneficial for CCSN searches.
Indeed, both surveys propose to image an area of 31000 deg$^2$, at $\sim$1.4 GHz,
reaching r.m.s. sensitivities of  2 (3) $\mu$Jy/beam after two years with an angular resolution of $\sim$0.5 ($\sim$2) arcsec resolution.
If any of those  surveys is performed, observing the cadence-time of $\sim$90 days, this would yield 
increments in the number of CCSNe expected to be detected with respect to the
nominal values in Table \ref{expectations} for SKA1-SUR by factors of 6 and 11, for the 
All-Sky surveys with 2$\mu$Jy/beam and 3$\mu$Jy/beam sensitivities.

Once SKA is completed, the sensitivity will be about 10 times better than for SKA1, the angular resolution will be about 20 times better, and the FoV around 20 times larger. Therefore, the number of detections is expected to be very close to the number of explosions of all CCSNe in the local universe (see Table~\ref{expectations}), and the close to milliarcsecond angular resolution could permit to detect CCSNe even in the dusty environments on (U)LIRGs, thus allowing us to obtain a dust-free, complete census of CCSNe in the local universe.
In addition, devoted CCSNe surveys of smaller fields (several degrees to a hundred) with sensitivities 
of $\sim 100$ nJy/beam (1 $\sigma$) will be possible with very modest observing times. This will allow us to precisely determine the value and evolution of the fundamental parameter $\Re$ as a function of redshift in the relevant redshift range from $z = 0$ to at $z \simeq 1.0$ from radio observations, and compare them with existing estimates (see right panel in Fig. \ref{missing-ccsne}).

\section{Type Ia supernovae with the SKA}
\label{sec:typeIa}

Type Ia SNe  are the end-products of white dwarfs  (WDs) with a
mass approaching, or equal to, the Chandrasekhar limit, which results in a
thermonuclear explosion of the star. 
While it is well acknowledged that  the exploding WD dies in close binary systems, 
it is still unclear whether the progenitor system is composed of a 
C+O white dwarf and a non-degenerate star (single-degenerate scenario), or 
both stars are WDs (double-degenerate scenario). 
In the single-degenerate scenario, a WD accretes mass from a
hydrogen-rich  companion star before reaching a mass close to the
Chandrasekhar mass and going off as supernova, while in the double-degenerate scenario, two
WDs merge, with the more-massive WD being thought to tidally disrupt and accrete the lower-mass WD \citep[see, e.g.,][and references therein]{maoz13}.
This lack of knowledge makes it difficult to
gain a physical understanding of the explosions, and to model their
evolution, thus also compromising their use as distance indicators. 

\subsection{Radio and X-ray observations of SNe Ia}
\label{sec:typeIa-obsns}

Radio and X-ray observations can potentially discriminate between the progenitor
models of SNe Ia.  For example, in all scenarios with mass
transfer from a companion, a significant amount of circumstellar gas
is expected \citep[see, e.g.,][]{branch95}, and therefore a shock is
bound to form when the supernova ejecta are expelled. The situation
would then be very similar to circumstellar interaction in
core-collapse SNe (see above), where the interaction of the blast wave from the
supernova with its circumstellar medium results in strong radio and
X-ray emission \citep{chevalier82}.  On the other hand, the
double-degenerate scenario will not give rise to any circumstellar medium close to the progenitor system,
and hence essentially no prompt radio emission is expected. 
Nonetheless, we note that the radio emission increases with time in the double-degenerate scenario, contrary to the single-degenerate scenario. This also opens the possibility for confirming the double-degenerate channel
in Type Ia SNe via sensitive SKA observations of decades-old Type Ia SNe.

\begin{figure}[htb]
\centering
\includegraphics[width=1.1\textwidth]{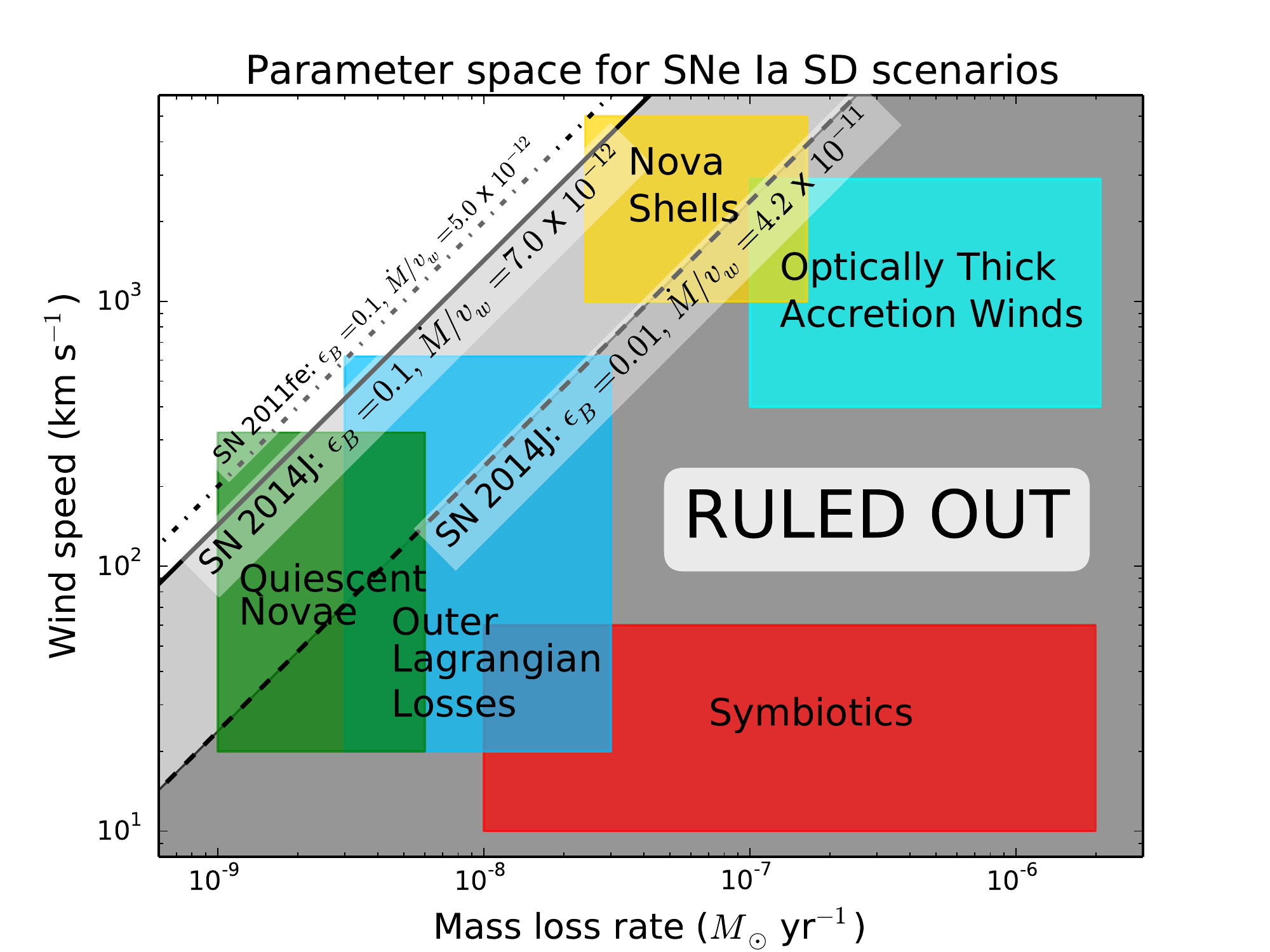}
\caption{\em Constraints on the parameter space (wind speed vs. mass-loss rate) for single degenerate scenarios for SN~2014J (see \citealt{mpt14} for details). Progenitor scenarios are plotted as schematic zones, following \citet{chomiuk12}. We indicate our 3$\sigma$ limits on $\mdot/v_{w}$, assuming $\epsilon_{\rm B}$ = 0.1 
(solid line) and the conservative case of $\epsilon_{\rm B}$ = 0.01 (dashed line). Mass loss scenarios falling into the gray-shaded areas should have been detected by the deep radio observations, and therefore are ruled out for SN~2014J. For a comparison, we have included also the limit on SN~2011fe (dash-dotted line) for the same choice of parameters as the solid line for SN~2014J, which essentially leaves only room for quiescent nova emission as a viable alternative among the single-degenerate scenarios for SN~2011fe (see \citet{mpt14} for details).
}
\label{mdot-vwind}
\end{figure}

Radio \citep[e.g.,][]{panagia06,hancock11} 
and X-ray \citep[e.g.,][]{hughes07,russell12} observations of SNe~Ia resulted
in upper limits on the wind density around SN Ia progenitors of the order of  
$\mdot = 1.2\times10^{-7}$ \msunyr, assuming a wind velocity of 10~km~s$^{-1}$.  
At the moment, the deepest radio limits on circumstellar gas come from SNe~2011fe and 2014J.
The limits on mass-loss rate from the progenitor system of SN~2011fe are $\mdot = 6\times10^{-10}$ \msunyr\ 
and $\mdot = 2\times10^{-9}$ \msunyr\ from 
radio \citep{chomiuk12} and X-rays \citep{margutti12}, respectively, assuming 
a wind velocity omass-lossf 100~km~s$^{-1}$.  
Similarly, the mass-loss rate limits for the progenitor system of SN~2014J are 
$\mdot = 7\times10^{-10}$ \msunyr\ 
and $\mdot = 1.2\times10^{-9}$ \msunyr\ from 
radio \citep{mpt14} and X-rays \citep{margutti14}, respectively, for 
a wind velocity of 100~km~s$^{-1}$.  
The above limits permit to rule out all symbiotic systems and the majority of the parameter space associated with stable nuclear burning WDs, as viable progenitor systems for either SN~2011fe or SN~2014J. Recurrent novae with main sequence or subgiant donors cannot be ruled out completely, yet most of their parameter space is also excluded by those radio observations (see Fig. \ref{mdot-vwind}.)

\begin{figure}[htb]
\centering
\includegraphics[width=1.1\textwidth]{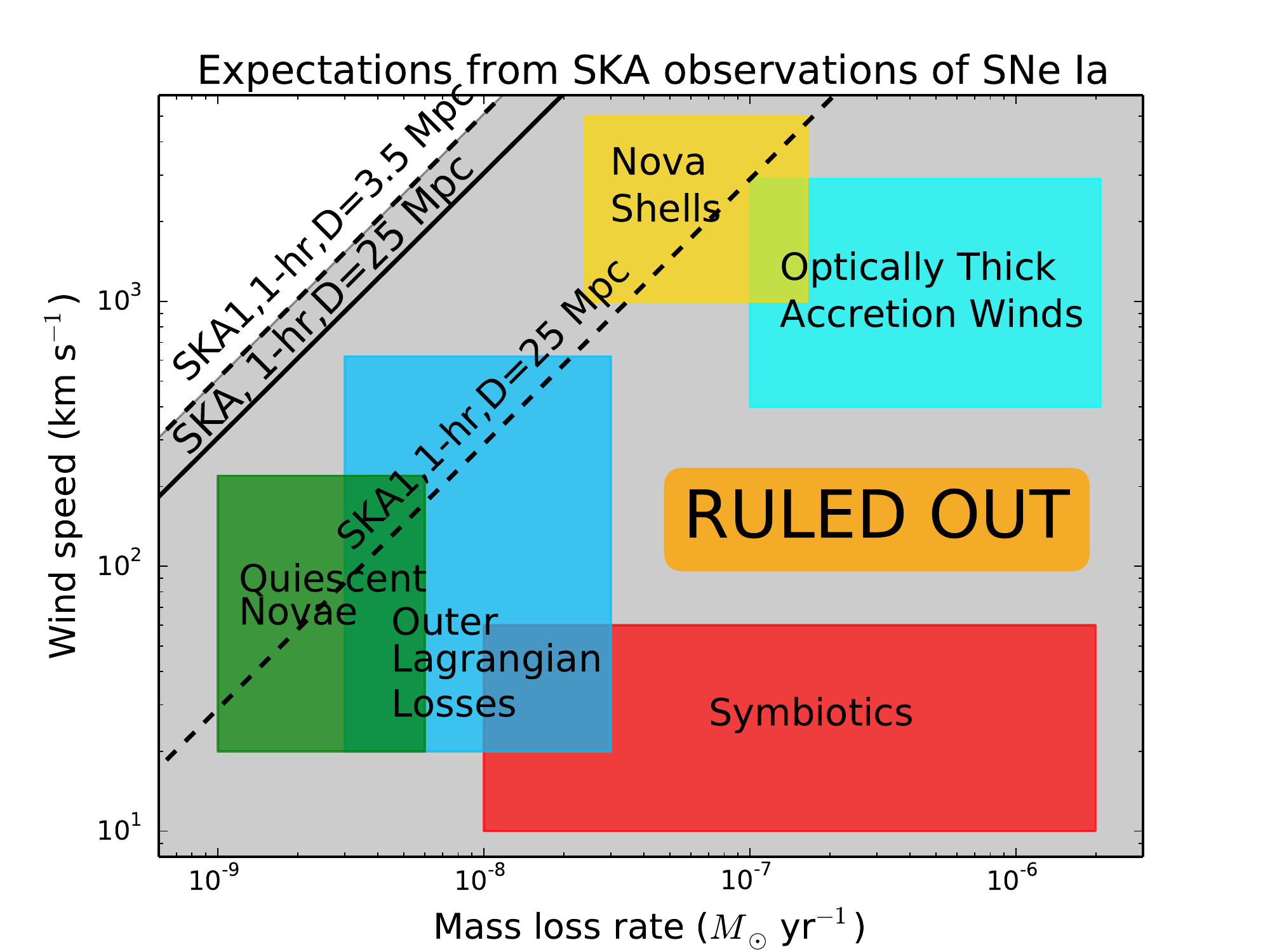}
\caption{\em Constraints on the parameter space (wind speed vs. mass-loss rate) for the same
single-degenerate scenarios as in Fig. 3, as expected with SKA1 (dashed lines) and the full SKA (solid line). SKA1 will be able to unambiguously probe all single degenerate scenarios for SNe Ia exploding at distances similar to that of M~82 (3.5 Mpc), and will be more sensitive than current state of the art, deep radio observations of SN~2014J in M~82, up to a distance of 25 Mpc, or even larger. When SKA is completed, we will be able to unambiguously probe the prompt radio emission within the single degenerate scenario up to distances of $\geq 25$ Mpc. All lines correspond to 3-$\sigma$.}
\label{fig:SKA}
\end{figure}

\subsection{Unveiling the progenitor scenarios of type Ia SNe with the SKA}
\label{sec:unveiling-typeIa}

 With the advent of the SKA, we will be able to obtain significantly deeper radio  limits, or an eventual detection, for SNe Ia exploding at the distance of M~82.  For more distant supernovae, we will be able to obtain similar or even more constraining limits to those obtained for SNe~2011fe and 2014J, which will allow us to build a picture from a larger statistical sample of observed SNe Ia.

The first phase of SKA considers three different surveys: SKA1-LOW, -mid, and -sur (see also Table 2). SKA1-MID, promises to yield 1$\sigma$ sensitivities of $\sim 0.7 \mu$Jy/beam in one hour at a fiducial frequency of 1.7 GHz. 
This figure is five times better than currently provided by the most sensitive array, the VLA. 
SKA1-MID will be able to either detect the putative radio emission of SN~2014J-like objects up to distances $\lesssim 8$ Mpc in less than one hour, or put significantly better constraints on the parameter space of single-degenerate scenarios for the next Type Ia SN that explodes at a distance no larger than that to M~82. 
However, the expected number of SNe Ia per year in such a volume of the local universe is small. 
Since the volumetric rate of Type Ia SNe is $\sim 3\times 10^{-5}$ SN\,yr$^{-1}$\,Mpc$^{-3}$ \citep{dilday10}, 
we should expect on average one Type Ia SN every $\sim$15 yr within a distance of $\lsim$8 Mpc (more than twice the distance to M~82), which is a depressingly small value to get  any statistical improvement in a sensible amount of time. 
\citet{smartt09} found 26 Type Ia SNe out of 132 SNe from a 10.5 year-long survey
within 28 Mpc. This figure corresponds to about 1 SN Ia every 13 yr within a distance of 8 Mpc, in agreement with the value found by \citet{dilday10}.

To obtain a statistically significant sample of SNe~Ia observed in radio and with similar
upper limits to those obtained for SNe 2011fe and 2014J, we need to sample significantly larger volumes and need
much more sensitive radio observations. For example,  by sampling out to a distance of 25 Mpc, we can 
expect $\sim$2 SNe~Ia per year within this sampled volume, which in 10 years would result in a total of $\sim$20 SNe~Ia, enough to extract statistical results.  
At this maximum distance, we need a sensitivity  $\sim$50 times better
than obtained by the deep radio observations discussed in \citet{chomiuk12} and \citet{mpt14}, 
i.e., 80 nJy/beam (1-$\sigma$), to be as constraining.
The fiducial 1-$\sigma$ sensitivity of SKA should be 10 times better than that of SKA1-MID, 
or about $\sim 70$~nJy/beam in one hour, which will allow to obtain deep radio limits (or eventually the detection) of type Ia SNe in a short amount of time and out to 25 Mpc, or even further. 
To get a more clear idea of what can be reached with SKA1 and SKA, we plot in Fig. \ref{fig:SKA}
the constraints on the mass-loss rate parameter for an upper limit of 
 3-$\sigma$  (3$\times 0.7\mu$Jy/beam for SKA1, or 3$\times$70 nJy/beam for SKA) 
 for a Type Ia SN exploding at the distance of M~82 and at 25 Mpc.
It is evident that, at this level of sensitivity, a non-detection would be essentially as meaningful as a  detection, since the former would imply that only the double-degenerate scenario is viable, while the latter would tell us which of the single-degenerate channels results in Type Ia SNe.
The overall time needed to carry out such a target-of-opportunity programme will require no more than about 12$-$24 hr/year, overheads included, for an average of two targets/year within a radius of 25 Mpc.
Such modest time requests can be easily accommodated within a sensible period of time.

\section{Summary}
\label{sec:summary}

We have presented the prospects for advancing our understanding of the physics of supernovae  via their study at radio wavelengths with the SKA. Our suggested approach for core-collapse supernovae is a commensal one, taking advantage of the deep, sensitive surveys that are planned with SKA1. 
We have discussed the expectations for CCSN studies under the specific case of 
SKA1-SUR ($\sim$1000 hr in one year, rms=4.2 $\mu$Jy/beam in 1-hr, FoV=18 deg$^2$, bandwidth=500 MHz). 
In particular, we assumed a fiducial frequency of 1.7 GHz and a covered area of 10000 deg$^2$ in one year.  
The expected number of new CCSNe discovered after one year would be $\gsim$310.
For SKA, the expected number of expected CCSN discoveries is well over 10,000 in one year.
Therefore, the number of detections is expected to approach the number of explosions of all CCSNe in the local universe, thus allowing us to obtain a dust-free, complete census of CCSNe.
The only request from such a programme is a multi-epoch approach, observing 
a cadence-time of $\lsim 90 \nu_{1.7}^{-1}\,$  days, where $\nu_{1.7} = \nu / 1.7$ GHz.
We also note that the proposed programme can be carried out as well at higher frequency bands, as long as an adequate cadence time is observed.

We have also discussed the prospects for probing Type Ia SNe progenitor scenarios with the SKA. 
The SKA can be used at very low time-cost for searching the putative prompt radio emission arising, in the single-degenerate scenario,
from the circumstellar medium around Type Ia SN progenitors in the nearby universe.
Complementarily, since the radio emission of Type Ia SNe is expected to increase with time in the double-degenerate scenario, the SKA should observe decades-old, nearby Type Ia SNe, which can potentially 
confirm the double-degenerate scenario in them.
In conclusion, the huge improvement in sensitivity of the SKA with respect to their predecessors should allow us to unambiguously discern which  progenitor scenario (single-degenerate vs. double-degenerate) applies to them, thus solving this long-standing issue.

\section{Acknowledgments}
\noindent
MAPT, AA, RHI, JMM, and ER acknowledge support from the Spanish Ministry of 
Economy and Competitiveness (MINECO) through grants AYA2012-38491-C02-01 and AYA2012-38491-C02-02.
PL acknowledges support from the Swedish Research Council.
JMM and ER also acknowledge support from the Generalitat Valenciana through grants PROMETEO/2009/104 and PROMETEOII/2014/057.
CRC acknowledges financial support from the ALMA-CONICYT FUND Project 31100004.
CRC is also supported by the Ministry of Economy, Development, and Tourism's Millennium Science Initiative through grant IC120009, awarded to The Millennium Institute of Astrophysics, MAS.


\bibliographystyle{plainnat}

\begin{thebibliography}{99}
\bibitem[Alberdi et al.(2006)]{alberdi06} Alberdi, A., Colina, 
L., Torrelles, J.~M., et al.\ 2006, \apj, 638, 938 
\bibitem[Batejat et al.(2011)]{batejat11} Batejat, F., Conway, 
J.~E., Hurley, R., et al.\ 2011, \apj, 740, 95 
\bibitem[Bietenholz et al.(2014)]{bietenholz14} Bietenholz, M.~F., 
De Colle, F., Granot, J., Bartel, N., 
\& Soderberg, A.~M.\ 2014, \mnras, 440, 821 
\bibitem[Bondi et 
al.(2012)]{bondi12} Bondi, M., P{\'e}rez-Torres, M.~A., Herrero-Illana, R., \& Alberdi, A.\ 2012, \aap, 539, A134 
\bibitem[Branch et al.(1995)]{branch95} Branch, D., Livio, M., 
Yungelson, L.~R., Boffi, F.~R., \& Baron, E.\ 1995, \pasp, 107, 1019 
\bibitem[Chevalier(1982)]{chevalier82} Chevalier, R.~A.\ 1982, 
\apj, 259, 302 
\bibitem[Chevalier(1998)]{chevalier98} Chevalier, R.~A.\ 1998, 
\apj, 499, 810 
\bibitem[Chevalier et al.(2006)]{chevalier06} Chevalier, R.~A., 
Fransson, C., \& Nymark, T.~K.\ 2006, \apj, 641, 1029 
\bibitem[Chomiuk et al.(2012)]{chomiuk12} Chomiuk, L., Soderberg, 
A.~M., Moe, M., et al.\ 2012, \apj, 750, 164 
\bibitem[Colina et al.(2001)]{colina01} Colina, L., Alberdi, A., 
Torrelles, J.~M., Panagia, N., \& Wilson, A.~S.\ 2001, \apjl, 553, L19 
\bibitem[Dahlen et al.(2012)]{dahlen12} Dahlen, T., Strolger, 
L.-G., Riess, A.~G., et al.\ 2012, \apj, 757, 70 
\bibitem[Dilday et al.(2010)]{dilday10} Dilday, B., Smith, M., 
Bassett, B., et al.\ 2010, \apj, 713, 1026 
\bibitem[Eldridge et al.(2013)]{eldridge13} Eldridge, J.~J., 
Fraser, M., Smartt, S.~J., Maund, J.~R., 
\& Crockett, R.~M.\ 2013, \mnras, 436, 774 
\bibitem[Gal-Yam et al.(2006)]{gal-yam06} Gal-Yam, A., Ofek, 
E.~O., Poznanski, D., et al.\ 2006, \apj, 639, 331 
\bibitem[Hancock et al.(2011)]{hancock11} Hancock, P.~J., 
Gaensler, B.~M., \& Murphy, T.\ 2011, \apjl, 735, L35 
\bibitem[Horiuchi et al.(2011)]{horiuchi11} Horiuchi, S., Beacom, 
J.~F., Kochanek, C.~S., et al.\ 2011, \apj, 738, 154 
\bibitem[Hughes et al.(2007)]{hughes07} Hughes, J.~P., Chugai, 
N., Chevalier, R., Lundqvist, P., \& Schlegel, E.\ 2007, \apj, 670, 1260 
\bibitem[Kamble et al.(2014)]{kamble14} Kamble, A., Soderberg, 
A., Berger, E., et al.\ 2014, arXiv:1401.1221 
\bibitem[Kankare et al.(2014)]{kankare14} Kankare, E., Mattila, 
S., Ryder, S., et al.\ 2014, \mnras, 440, 1052 
\bibitem[Lien et al.(2011)]{lien11} Lien, A., Chakraborty, N., 
Fields, B.~D., \& Kemball, A.\ 2011, \apj, 740, 23 
\bibitem[Magnelli et 
al.(2011)]{magnelli11} Magnelli, B., Elbaz, D., Chary, R.~R., et al.\ 2011, \aap, 528, A35 
\bibitem[Mannucci et al.(2007)]{mannucci07} Mannucci, F., Della 
Valle, M., \& Panagia, N.\ 2007, \mnras, 377, 1229 
\bibitem[Maoz et al.(2013)]{maoz13} Maoz, D., Mannucci, F., 
\& Nelemans, G.\ 2013, arXiv:1312.0628 
\bibitem[Margutti et al.(2012)]{margutti12} Margutti, R., 
Soderberg, A.~M., Chomiuk, L., et al.\ 2012, \apj, 751, 134 
\bibitem[Margutti et al.(2014)]{margutti14} Margutti, R., Parrent, 
J., Kamble, A., et al.\ 2014, \apj, 790, 52 
\bibitem[Mattila et al.(2012)]{mattila12} Mattila, S., Dahlen, 
T., Efstathiou, A., et al.\ 2012, \apj, 756, 111 
\bibitem[Matzner \& McKee(1999)]{matzner99} Matzner, C.~D., \& McKee, C.~F.\ 1999, \apj, 510, 379 
\bibitem[Panagia et al.(2006)]{panagia06} Panagia, N., Van Dyk, 
S.~D., Weiler, K.~W., et al.\ 2006, \apj, 646, 369 
\bibitem[Parra et al.(2007)]{parra07} Parra, R., Conway, J.~E., 
Diamond, P.~J., et al.\ 2007, \apj, 659, 314 
\bibitem[P{\'e}rez-Torres et al.(2009a)]{mpt09a} 
P{\'e}rez-Torres, M.~A., Alberdi, A., Colina, L., et al.\ 2009, \mnras, 
399, 1641 
\bibitem[P{\'e}rez-Torres et 
al.(2009b)]{mpt09b} P{\'e}rez-Torres, M.~A., Romero-Ca{\~n}izales, C., Alberdi, A., \& Polatidis, A.\ 2009, \aap, 507, L17 
\bibitem[P{\'e}rez-Torres et 
al.(2010)]{mpt10} P{\'e}rez-Torres, M.~A., Alberdi, A., Romero-Ca{\~n}izales, C., \& Bondi, M.\ 2010, \aap, 519, L5 
\bibitem[P{\'e}rez-Torres et al.(2014)]{mpt14} 
P{\'e}rez-Torres, M.~A., Lundqvist, P., Beswick, R.~J., et al.\ 2014, \apj, 
792, 38 
\bibitem[Romero-Ca{\~n}izales et al.(2011)]{crc11} 
Romero-Ca{\~n}izales, C., Mattila, S., Alberdi, A., et al.\ 2011, \mnras, 
415, 2688 
\bibitem[Romero-Ca{\~n}izales et al.(2014)]{crc14} 
Romero-Ca{\~n}izales, C., Herrero-Illana, R., P{\'e}rez-Torres, M.~A., et 
al.\ 2014, \mnras, 440, 1067 
\bibitem[Russell \& Immler(2012)]{russell12} Russell, B.~R., \& Immler, S.\ 2012, \apjl, 748, L29 
\bibitem[Smartt(2009)]{smartt09} Smartt, S.~J.\ 2009, \araa, 47, 63 
\bibitem[Soderberg et al.(2006)]{soderberg06} Soderberg, A.~M., 
Nakar, E., Berger, E., \& Kulkarni, S.~R.\ 2006, \apj, 638, 930 
\bibitem[Weiler et 
al.(2002)]{weiler02} Weiler, K.~W., Panagia, N., Montes, M.~J., \& Sramek, R.~A.\ 2002, \araa, 40, 387 
\end{thebibliography}

\end{document}